\documentclass[aps,pre,twocolumn]{revtex4-2}
\usepackage{amsmath,bm,graphicx}

\begin{document}
\title{Dilute Limit Coarsening with an Anisotropic Surface Tension}

\author{Arjun R. Anand}
\author{Melinda M. Andrews}
\author{Benjamin P. Vollmayr-Lee}
\email{bvollmay@bucknell.edu}
\affiliation{Department of Physics \&\ Astronomy, 
  Bucknell University, Lewisburg, Pennsylvania 17837, USA}

\date{\today}

\begin{abstract}
We investigate the impact of an anisotropic surface tension on the
late-stage dilute phase separation dynamics, revisiting the seminal
Lifshitz-Slyozov (LS) theory, which traditionally relies on the
assumption of isotropic surface tension. Using a perturbative
treatment for weak anisotropy, we demonstrate that although the
characteristic $t^{1/3}$ drop growth law remains unchanged, the
anisotropy causes a significant breakdown of morphological
universality. Specifically, we calculate explicitly a one-parameter
family of nonequilibrium drop shapes that depend on the scaled drop
size.  These shapes are close to the equilibrium Wulff shape, but the
smaller drops are more spherical and the larger drops have an enhanced
anisotropy in comparison to the Wulff shape. We also demonstrate that
the the drop size distribution is modified from the isotropic LS
distribution at second order in the anisotropy strength.
\end{abstract}

\maketitle

\section{Introduction}

\label{sec:introduction}

Coarsening, or phase separation dynamics, constitutes an important
category of nonequilibrium phenomena in which domains of distinct
phases grow over time, generally driven by the reduction of
interfacial energy (see \cite{Gunton1983,Bray1994} for reviews). Such
behavior is ubiquitous, appearing in a wide range of physical and
biological systems, including binary alloys
\cite{Cahn1994,Orlikowski1999}, foams
\cite{Stavans1993,Hilgenfeldt2001b,Hilgenfeldt2001,Thomas2006,Lambert2010},
magnetic systems \cite{Singh2023}, polymer blends
\cite{Chakrabarti1990,Wang2000}, and even cellular aggregates
\cite{Weyer2024} and Bose superfluids \cite{Gliott2025}. Common
features observed across these systems are self-similarity, power-law
growth, and a degree of universality. A major goal in the theoretical
study of coarsening is to elucidate and understand these
characteristic features.

A landmark result in the theoretical study of thermodynamic coarsening
is Lifshitz-Slyozov (LS) theory \cite{Lifshitz1961}, which provides an
exact solution for the late-stage dynamics of dilute phase separation
with a conserved scalar order parameter. In this dilute limit, an
isolated drop
or domain of one phase is immersed a matrix of the other phase and
interacts with other drops only through the amount of
supersaturation in the matrix.  This framework predicts two
characteristic universal features: a $t^{1/3}$ drop growth law for the
characteristic drop size and an explicit universal scaling function
for the drop size distribution.  LS theory was extended by Wagner to
the case of a globally conserved order parameter, for which the growth
law becomes $t^{1/2}$
 \cite{Wagner1961}.

Despite decades of study, our understanding of coarsening in passive
systems still contains gaps, partly due to the limited number of exact
results available. Indeed, to this day, LS theory represents the only
exact solution for subcritical $T<T_c$ coarsening.  While LS theory is
restricted to the dilute limit, it has proven remarkably flexible,
having been adapted successfully to study factors such as finite-size
effects \cite{Heermann1996}, mobility constraints \cite{Bray1995}, and
persistence properties \cite{Lee1997,Soriano2009}. However, these
extensions have largely relied on the critical assumption of an
isotropic surface tension.

One question that has not been fully resolved is the impact of an
anisotropic surface tension on coarsening structure \cite{Siegert1990,
  McFadden1993,Rutenberg1996}.  In realistic systems, especially those
involving crystal growth, the surface tension is often
orientation-dependent, dictating that the equilibrium shape of a
precipitate --- determined by the Wulff construction --- is
non-spherical \cite{Wulff1901,Herring1953,Wortis1988}. This shape
perturbation is expected to influence the diffusive mass transport
surrounding the particle and, consequently, alter the overall kinetics
of the coarsening process, potentially affecting domain morphology and
universality.  

In this work, we revisit LS theory to study coarsening with
anisotropic surface tension. One central question is whether the
domain shapes are equilibrium Wulff shapes, a new nonequilibrium
shape, or perhaps a family of drop shapes dependent on the scaled
size of the drops.  A second question is whether the drop size distribution is
universal or instead sensitive to details of the anisotropy. Lifshitz
and Slyozov state in a footnote their expectation that drops (or
grains) ``grow in such a way as to preserve their shape due to an
anisotropy in [the surface tension]''and further state in their
conclusion that the `effects of anisotropy and of the nonspherical
nature of the original grain nuclei \dots can all be taken into
account simply by the use of ``effective'' parameters in place of
certain quantities in the expression for the critical size.  The form
of the distribution function is unaffected by these considerations.'
\cite{Lifshitz1961}.

Contrary to these claims, we find via a perturbative treatment for
weak anisotropy that surface tension anisotropy leads to a
one-parameter family of nonequilibrium drop shapes dependent on scaled
size.  We also demonstrate that the size distribution is affected
by anisotropy, indicating that the morphological universality
breaks down even though the growth law remains unchanged.  Similar
results were obtained in a earlier study of anisotropic coarsening of
a globally conserved order parameter in the dilute limit, building on
Wagner theory \cite{Rutenberg1999}, but that calculation differs
significantly from the current study of a locally conserved order
parameter due to the lack of a diffusion field.

Our technique is to begin with a generic, weak surface tension anisotropy
characterized by an expansion in spherical harmonics:
\begin{equation}
  \sigma(\theta_n,\phi_n) = \sigma_0\biggl[ 1 + \delta \sum_{\ell>0,m} s_{\ell m}
    Y_\ell^m(\theta_n,\phi_n) \biggr].
  \label{eq:sigma-expansion}
\end{equation}
Here $\theta_n$ and $\phi_n$ are the polar and azimuthal angles of the
interface unit normal $\hat{\bf n}$, while the sum is over $\ell=0, 1, \dots$ and $m = -\ell, \dots \ell$.  We introduce $\delta$ as a bookkeeping parameter
for the order of perturbation theory, though one could equivalently set
$\delta=1$ and organize the calculation in powers of the $s_{\ell m}$
coefficients.
We then parameterize an arbitrary anisotropic
drop shape similarly,
\begin{equation}
  R(\theta,\phi) = R_0\biggl[ 1 + \sum_{\ell,m} (\delta a_{\ell m}
    + \delta^2 b_{\ell m} + \dots) Y_\ell^m(\theta,\phi) \biggr]
  \label{eq:R-expansion}
\end{equation}
where $\theta$ and $\phi$ are simply spherical coordinates of
three-dimensional space, i.e., the equation $r=R(\theta,\phi)$ defines
the drop surface.  We derive the equation of motion for this drop for
not just the size $R_0$ but also for the shape parameters $a_{\ell
  m}$, $b_{\ell m}$, \dots, and the look for a scaling solution.  In
general in this scaling solution there will be higher order
corrections to the drop shape, necessitating the expansion in powers
of $\delta$.  To organize the calculation, we make the arbitrary but
convenient choice to demand that $R_0$ evolves according to the
isotropic LS theory.  This choice requires that we include
the $\ell=0$ terms in Eq.~(\ref{eq:R-expansion}) to account for 
the impact on anisotropy on the drop size.

Our primary results are (1) a demonstration that to first order the
shape coefficients $a_{\ell m}$ have a scaling form and provide an
explicit solution for the one-parameter family of nonequilibrium drop
shapes, and (2) we provide evidence that the drop size distribution is
modified at second order in anisotropy strength.

The remainder of the paper is as follows.  In
Sec.~\ref{sec:preliminaries} we review the phenomenology of late-stage
coarsening dynamics and also review the isotropic LS theory, to
provide a reference for what follows.  Then in
Sec.~\ref{sec:anisotropic-gibbs-thomson} we derive the anisotropic
Gibbs-Thomson condition for the chemical potential at the drop surface
and evaluate it in terms of our parametrizations in
Eqs.~(\ref{eq:sigma-expansion}) and (\ref{eq:R-expansion}) to first
order in anisotropy strength.  In
Sec.~\ref{sec:equilibrium-drop-shape} we review the equilibrium Wulff
shape, which provides a useful comparison for our results.  We then
solve for the chemical potential inside and outside of the drop and
derive the resulting interface dynamics in
Sec.~\ref{sec:interface-dynamics}.  Then we turn to the question of
drop center motion and work out how this impacts the evolution of the
drop shape function $R(\theta,\phi)$ in
Sec.~\ref{sec:drop-center-motion}.  At this point we will have the
ingredients to solve for the scaling form of the drop shape parameters
$a_{\ell m}$ which we present in Sec.~\ref{sec:scaling-solution},
providing the main result of this paper.  In
Sec.~\ref{sec:drop-size-distribution} we develop the implications for
the drop size distribution, and finally in Sec.~\ref{sec:summary} we
summarize our results.

\section{Preliminaries}

\label{sec:preliminaries}

We begin with reviewing both the phenomenology of late-stage
conserved order parameter coarsening dynamics and isotropic
Lifshitz-Slyozov theory \cite{Lifshitz1961}, both for completeness and
to set up the framework to generalize Lifshitz-Slyozov theory to the
case of an anisotropic surface tension.

\subsection{Coarsening Phenomenology}

\label{subsec:coarsening-phenomenology}

Our treatment follows that of Bray \cite{Bray1994}.  Consider a binary
mixture, not necessarily in the dilute limit, of two components $A$
and $B$ with a uniform \textit{total} concentration.  Take
$\rho({\bf r})$ to be the local concentration of component $A$.  In
the late stages of coarsening, near-equilibrium domains are of size
$L(t)\sim t^{1/3}$, separated by interfacial regions of width $\xi$.
Within the domains the concentration field will be very close to one
of the equilibrium values, $\rho_1$ and $\rho_2$, corresponding to the
$A$-rich and $B$-rich phases, respectively.  Note that
$\rho_1>\rho_2$.

A given inhomogeneous concentration $\rho({\bf r})$ will have
associated with it a free energy $F[\rho({\bf r})]$, from which
the local chemical potential is defined via
\begin{equation}
  \mu({\bf r}) \equiv \frac{\delta F}{\delta\rho({\bf r})},
  \label{eq:chemical-potential-def}
\end{equation}
i.e., $\mu({\bf r})$ describes the response of the free energy to a
small, local concentration change.  A common phenomenological choice
for the free energy functional corresponding to an \textit{isotropic}
surface tension is
\begin{equation}
  F[\rho({\bf r})] = \int d{\bf r}\, \biggl\{\frac{1}{2}\kappa^2
  (\nabla\rho)^2 + V(\rho) \biggr\}.
  \label{eq:isotropic-free-energy}
\end{equation}
where $V(\rho)$ is a double-well potential.  For this choice one
obtains from Eq.~(\ref{eq:chemical-potential-def}) $\mu = V'(\rho) -
\kappa^2\nabla^2\rho$.  The local chemical potential plays a important
role both in defining our phenomenological dynamics and in solving the
resulting theory.

Because of the conserved dynamics, the density $\rho$ obeys a
continuity equation $\partial_t\rho = - \nabla\cdot{\bf J}$, where the
current (of component $A$) is presumed to be driven by gradients in
the chemical potential.  At sufficiently late stages of coarsening, as
the domains become large, spatial variations in the chemical potential
will becoming increasingly gradual, justifying the asymptotic relation
\begin{equation}
  {\bf J} = - M \nabla \mu.
  \label{eq:current}
\end{equation}

Within the nearly equilibrated domains, which we will refer to henceforth
as the bulk region (i.e.,\ not near an interface), the
concentration can be written as $\rho({\bf r}) = \rho_\text{eq} +
\delta\rho({\bf r})$, where $\rho_\text{eq}$ is either $\rho_1$ or $\rho_2$
and $\delta\rho({\bf r})$ is a small
supersaturation.  Plugging this into the free energy functional and
expanding $\delta\rho$ to first order results in
\begin{equation}
  \mu({\bf r}) = V''(\rho_\text{eq}) \, \delta\rho({\bf r}),
\end{equation}
i.e., the local chemical potential is directly proportional to the
supersaturation field.  Combining this with Eq.~(\ref{eq:current}) and
the continuity equation results in the diffusion equation
\begin{equation}
  \partial_t\mu = M V''(\rho_\text{eq}) \nabla^2\mu
\end{equation}
within the bulk regions.

In late-stage coarsening, when the characteristic length scale $L$ is
large compared to other lengths in the system, the characteristic
interface velocity scales as $v\sim \dot L \sim L^{-2}$.  The
diffusion of the bulk chemical potential equilibrates on time scale
$\tau \sim L^2$, during which time the interfaces will move a
distance $v\tau\sim L^{0}$.  This motion is negligible for a
domain of size $L$, thus we are justified in making the
quasistatic approximation that
\begin{equation}
  \nabla^2\mu = 0
  \label{eq:laplaces-eq}
\end{equation}
in the bulk regions.

Further, since the domain size $L(t)$ is asymptotically much larger
than the fixed interface width $\xi$, we may take the \textit{sharp-interface
limit}.  In this limit the chemical potential at the interface is given
by the Gibbs-Thomson relation (derived below).  Since the chemical
potential is thus prescribed along the interface and otherwise satisfies
Laplace's equation, this provides a Dirichlet problem, the solution of
which specifies the chemical potential everywhere.

Finally, we use knowledge of this chemical potential solution to determine
the interface motion.  Due to the conserved dynamics, an interface can
only move when there is a discontinuity in the normal component of the
current across the interface.  From Eq.~(\ref{eq:current}) we
obtain the normal interface velocity from
\begin{equation}
  v_n({\bf r}) = \frac{M}{\Delta\rho} \hat{\bf n}\cdot \lim_{\epsilon\to 0}
  [\nabla\mu({\bf r}+\epsilon\hat{\bf n}) - \nabla\mu({\bf r}-\epsilon
    \hat{\bf n})]
  \label{eq:velocity}
\end{equation}
where $\Delta\rho = \rho_1-\rho_2 > 0$ and we take the interface
normal $\hat{\bf n}$ to point away from the $A$-rich phase.
Eq.~(\ref{eq:velocity}) fully specify the asymptotic dynamics of
the interfaces.

\subsection{Isotropic Lifshitz-Slyozov Theory}

\label{eq:ls-theory}

In the dilute limit we may consider a single, isolated spherical drop
of $A$-rich phase within a matrix of $B$-rich phase.  This drop will
interact with other drops only through the supersaturation field at
large $r$.  Let $R_0$ be the drop radius (we use $R_0$ for consistency
with later notation).  The free energy of this drop
is simply the surface area times the isotropic surface tension
$\sigma$, i.e.,\ $F=4\pi R_0^2\sigma$.  To obtain the chemical potential
at the surface, presumed to be uniform, consider the variation $R_0\to
R_0+\delta R$, which will result in a free energy change $\delta F = \mu
\Delta\rho \delta V$.  Solving for $\mu$ provides the Gibbs-Thomson
result for a spherical drop with an isotropic surface tension,
\begin{equation}
  \mu(r=R_0) = \frac{2\sigma}{\Delta\rho} \frac{1}{R_0}.
  \label{eq:mu-gt-isotropic}
\end{equation}

Solving Laplace's equation with this boundary condition gives
\begin{equation}
  \mu({\bf r}) =
    \begin{cases} \frac{2\sigma}{\Delta\rho} \frac{1}{R_0} & r < R_0 \\
    \mu_\infty + (\frac{2\sigma}{\Delta\rho} - \mu_\infty R_0)\frac{1}{r}
    & r>R_0.
    \end{cases}
    \label{eq:ls-mu}
\end{equation}
where $\mu_\infty$ is related to the amount of supersaturation far
from the drop, and is to be determined.  The normal velocity of the
interface is in the radial direction, so using (\ref{eq:ls-mu}) in
(\ref{eq:velocity}) gives
\begin{equation}
  v_r = \frac{\alpha}{R_0}\biggl(\frac{1}{R_c} - \frac{1}{R_0}
  \biggr),
  \label{eq:ls-drop-dynamics}
\end{equation}
where $\alpha = 2M\sigma/(\Delta\rho)^2$ and
$R_c = 2\sigma/(\Delta\rho\mu_\infty)$ is the critical radius: drops
larger than $R_c$ will grow, while drops smaller than $R_c$ will shrink.
In general, $R_c$ may depend on time, if the supersaturation far from the
drops is time-varying.

The number density of drops of size $R_0$ is given by $n(R_0,t)$, which obeys
a continuity equation
\begin{equation}
  \frac{\partial n}{\partial t} = -\frac{\partial}{\partial R_0}(v_r n).
  \label{eq:ls-continuity}
\end{equation}
Following LS, we assume this distribution has the scaling form
\begin{equation}
  n(R_0,t) = \frac{1}{R_c(t)^4} f\Bigl(R_0/R_c(t)\Bigr)
\end{equation}
where the scaling prefactor comes from the requirement that $A$-rich
phase occupies a fixed fraction $\epsilon \ll 1$ of the spatial volume:
\begin{equation}
  \int_0^{\infty} dR_0 \,\frac{4}{3}\pi R_0^3\, n(R_0,t)
  = \frac{4\pi}{3}\int_0^\infty dx\,x^3 f(x) = \epsilon
\end{equation}
where we have introduced the scaled drop size $x=R_0/R_c(t)$.

Combining (\ref{eq:ls-drop-dynamics}) and (\ref{eq:ls-continuity}) we
find a scaling solution to be obtained provided that
\begin{equation}
  R_c = \biggl(\frac{4}{9} \alpha t\biggr)^{1/3}
  \label{eq:Rc}
\end{equation}
with the resulting equation for the distribution
\begin{equation}
  x(x+3)(\frac{3}{2}-x)^2 f'(x) + \frac{27}{4}(2-x-\frac{16}{27} x^3) f(x)
  = 0.
\end{equation}
This has the solution
\begin{equation}
  f(x) = A\epsilon \frac{x^2}{(x+3)^{7/3}(\frac{3}{2}-x)^{11/3}} \exp\biggl(
  -\frac{3}{3-2x}\biggr)
  \label{eq:ls-distribution}
\end{equation}
for the range $0<x<3/2$, with $A\simeq 14.6572$ determined from the
normalization integral.

\section{Anisotropic Gibbs-Thomson Relation}

\label{sec:anisotropic-gibbs-thomson}

We now consider how the Gibbs-Thomson relation is modified by having
both an anisotropic surface tension $\sigma(\hat{\bf n})$ that varies
with the interface normal $\hat{\bf n}$, and a non-spherical drop
shape.  A well-known result is that the chemical potential at a point
$P$ on the surface is given by
\begin{equation}
  \mu = \frac{1}{\Delta\rho}
  \biggl(\sigma + \frac{\partial^2\sigma}{\partial\theta_1^2}\biggr)
  \frac{1}{R_1} +
  \biggl(\sigma + \frac{\partial^2\sigma}{\partial\theta_2^2}\biggr)
  \frac{1}{R_2},
  \label{eq:anisotropic-gt-classic}
\end{equation}
where $\sigma$ and its derivatives are evaluated at ${\hat{\bf n}}$,
$R_1$ and $R_2$ are the principal radii of curvature at $P$, and
$\theta_1$ and $\theta_2$ are the angles with respect to $\hat{\bf n}$
within the corresponding principle planes
\cite{Herring1953,Johnson1965}.  We observe that in the case of an
isotropic surface tension but a non-spherical drop
Eq.~(\ref{eq:anisotropic-gt-classic}) reduces to the surface tension
times the mean curvature.

\begin{figure}
  \includegraphics[width=2.8in]{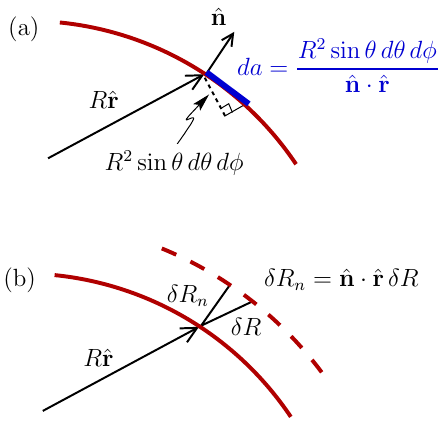}
  \caption{(a) The differential area of the anisotropic interface is
    simply related to the differential area of sphere of radius $R$ by
    a cosine factor given by $\hat{\bf n}\cdot\hat{\bf r}$.  (b) A
    displacement of the interface by an amount $\delta R$ in the
    radial direction results in a normal displacement of $\hat{\bf
      n}\cdot\hat{\bf r}\delta R$.}
  \label{fig:da}
\end{figure}

For our purposes, we assume a convex drop shape with a suitably chosen
origin (discussed in Sec.~\ref{sec:drop-center-motion}) such that we
can describe the shape in spherical coordinates by
\begin{equation}
  {\bf R} = R(\theta,\phi) \, \hat{\bf r}.
  \label{eq:drop-shape}
\end{equation}
While it is possible to proceed using (\ref{eq:anisotropic-gt-classic}),
it is more straightforward to re-derive the Gibbs-Thomson relation
from (\ref{eq:drop-shape}).  We first find the normal vector
\begin{equation}
   \hat{\bf n} = \frac{{\bf R}_\theta\times {\bf R}_\phi}{|{\bf R}_\theta
     \times {\bf R}_\phi|}
   =  \frac{\hat{\bf r} - \frac{R_\theta}{R}\hat{\bm\theta} -
     \frac{R_\phi}{R\sin\theta}
     \hat{\bm\phi}}{\Bigl[1 + \frac{R_\theta^2}{R^2}
        + \frac{R_\phi^2}{R^2\sin^2\theta}\Bigr]^{1/2}},
  \label{eq:normal}
\end{equation}
where ${\bf R}_i \equiv \partial_i {\bf R}$ and $R_i \equiv \partial_i
R$ for $i=\theta,\phi$.  The metric tensor $g_{ij}={\bf R}_i\cdot{\bf
  R}_j$ evaluates to
\begin{equation}
  g := \begin{pmatrix} R^2+R_\theta^2 & R_\theta R_\phi \\
    R_\theta R_\phi & R^2\sin^2\theta + R_\phi^2 \end{pmatrix}
\end{equation}
and provides the differential area
\begin{equation}
  da = \sqrt{g}\,d\theta\,d\phi = R^2\sin\theta \biggl[
    1 + \frac{R_\theta^2}{R^2}
    + \frac{R_\phi^2}{R^2\sin^2\theta}\biggr]^{1/2}\,d\theta\,d\phi.
\end{equation}
This has a simple geometric interpretation, as indicated in
Fig.~\ref{fig:da}a.

The free energy of this drop is given by $F=\int da\,\sigma(\hat{\bf
  n})$, or
\begin{equation}
  F 
  = \int d\Omega\, f(R,R_\theta,R_\phi;\theta,\phi).
  \label{eq:free-energy}
\end{equation}
where $d\Omega = \sin\theta\, d\theta\,d\phi$, the integrand is
\begin{equation}
  f(R,R_\theta,R_\phi;\theta,\phi)= \frac{R^2\sigma(\hat{\bf n})}
  {\hat{\bf n}\cdot\hat{\bf r}},
  \label{eq:f}
\end{equation}
and the integral is over the unit sphere.  The first order change in
free energy $\delta F$ due to a perturbation $R(\theta,\phi) \to
R(\theta,\phi) + \delta R(\theta,\phi)$ is
\begin{equation}
  \delta F = \int d\Omega\,\biggl[\frac{\partial f}{\partial R}
    - \frac{\partial}{\partial\theta}\frac{\partial f}{\partial R_\theta}
    - \cot\theta\frac{\partial f}{\partial R_\theta} - \frac{\partial}
    {\partial\phi} \frac{\partial f}{\partial R_\phi}\biggr]\delta R,
  \label{eq:deltaF-1}
\end{equation}
where we have used integration by parts and have assumed $f$ to be
finite at $\theta = 0$ and $\pi$, which eliminates the boundary terms.
Note that the partial derivatives with respect to $\theta$ and $\phi$
act on the implicit angular dependence in $R$, $R_\theta$, and
$R_\phi$.

We may separately express the free energy change due to a local
displacement as the product of the chemical potential, the volume that
changed phase, and the change in the field $\Delta\rho$.  As shown in
Fig.~\ref{fig:da}b, the volume change associated with the displacement
$\delta R$ is given by
\begin{equation}
  dV = da\, \delta R_n = da \, \hat{\bf n}\cdot \hat{\bf r}\,\delta R
  = d\Omega\, R^2 \,\delta R
\end{equation}
Thus the free energy change is
\begin{equation}
  \delta F = \int d\Omega\, \mu R^2\Delta\rho \,\delta R.
  \label{eq:deltaF-2}
\end{equation}
Equating (\ref{eq:deltaF-1}) and (\ref{eq:deltaF-2}) provides the
anisotropic Gibbs-Thomson result
\begin{equation}
  \mu = \frac{1}{\Delta\rho R^2}\biggl[\frac{\partial f}{\partial R}
    - \frac{\partial}{\partial\theta}\frac{\partial f}{\partial R_\theta}
    - \cot\theta\frac{\partial f}{\partial R_\theta} - \frac{\partial}
    {\partial\phi} \frac{\partial f}{\partial R_\phi}\biggr],
  \label{eq:anisotropic-gt}
\end{equation}
with $f$ given by Eq.~(\ref{eq:f}).  Some details of the evaluation of
(\ref{eq:anisotropic-gt}) are presented in Appendix
\ref{sec:appendix}.

The mean curvature for a drop with shape (\ref{eq:drop-shape}) has
been previous determined \cite{Mazharimousavi2017}, and we have
confirmed that in the simplified case of \textit{isotropic} surface
tension $\sigma_0$, our chemical potential (\ref{eq:anisotropic-gt})
indeed reduces to the mean curvature.

We now turn to the perturbative calculation to first order in the
drop shape and surface tension anisotropy.  This requires expressing
the normal angles, defined via $\hat{\bf n} = (\sin\theta_n\cos\phi_n,
\sin\theta_n\sin\phi_n,\cos\theta_n)$, in terms of the drop shape and
$\theta$ and $\phi$, as
\begin{equation}
  \theta_n = \arccos\biggl( \frac{\cos\theta + \frac{R_\theta}{R}\sin\theta}
        {[1+R_\theta^2/R^2 + R_\phi^2/(R^2\sin^2\theta)]^{1/2}}\biggr)
        \label{eq:thetan}
\end{equation}
and
\begin{equation}
  \phi_n = \phi - \arctan\biggl(\frac{R_\phi}{R\sin^2\theta -
    R_\theta\cos\theta\sin\phi}\biggr).
  \label{eq:phin}
\end{equation}
Expanding these to first order we obtain
\begin{equation}
  \mu = \frac{2\sigma_0}{\Delta\rho R_0}\biggl[
    1 + \delta\sum_{\ell,m} \beta_\ell (a_{\ell m}
    - s_{\ell m})Y_\ell^m(\Omega)\biggr] + O(\delta^2).
  \label{eq:mu-gt-first-order}
\end{equation}
with
\begin{equation}
  \beta_\ell \equiv \ell(\ell+1)/2 - 1
  \label{eq:beta}
\end{equation}
introduced for brevity.  This is our primary result for this section.

\section{Equilibruim Drop Shape}

\label{sec:equilibrium-drop-shape}

Before we proceed to the general dynamical equation of anisotropic drops,
we first turn to the equilibrium drop shapes as they provide a useful
frame of reference.

Given an anisotropic surface tension, the Wulff construction provides the
equilibrium shape, or Wulff shape, 
\begin{equation}
  R_\text{eq}(\hat{\bf r}) = \frac{R_0}{\sigma_0} \min_{\hat{\bf n}}
  \biggl(\frac{\sigma(\hat{\bf n})}{\hat{\bf n}\cdot\hat{\bf r}}\biggr),
  \label{eq:wulff-construction}
\end{equation}
which is the minimum energy shape corresponding to a fixed drop volume
\cite{Wulff1901,Herring1953,Wortis1988}.  Interestingly, the drop
shape and the surface tension as a function of the interface normal
provide a Legendre transform pair, with the inverse of the Wulff
construction, Eq.~(\ref{eq:wulff-construction}), given by the Herring
construction \cite{Herring1953,Wortis1988}
\begin{equation}
  \sigma(\hat{\bf n}) = \frac{\sigma_0}{R_0} \max_{\hat{\bf r}}
  \Bigl( R_\text{eq}(\hat{\bf r})\, \hat{\bf n}\cdot\hat{\bf r}\Bigr).
\end{equation}
Note that $R_0$ and $\sigma_0$ are parameters with dimension to scale the
drop and surface tension appropriately.

\subsection{The Wulff shape to first order}

The particular value of $\hat{\bf n}$ that minimizes the right hand
side of Eq.~(\ref{eq:wulff-construction}) for a given $\hat{\bf r}$
corresponds to the interface normal at that point.  Thus the Wulff
shapes obey
\begin{equation}
  \frac{R_\text{eq}(\theta,\phi)}{R_0} \biggl[ 1+\frac{R_\theta^2}{R^2}
   +\frac{R_\phi^2}{R^2\sin^2\theta}\biggr]^{-1/2} = \frac{\sigma(\theta_n,
    \phi_n)}{\sigma_0}.
\end{equation}
Expanding both sides to order $\delta$ gives
\begin{equation}
  1 + \delta\sum_{\ell m} a^\text{eq}_{\ell m} Y_\ell^m(\theta,\phi) 
  = 1 + \delta\sum_{\ell m} s_{\ell m} Y_\ell^m(\theta,\phi)
\end{equation}
where we have used $\theta_n=\theta + O(\delta)$ and $\phi_n=\phi+ O(\delta)$
from Eqs.~(\ref{eq:thetan}) and (\ref{eq:phin}).  From the orthogonality
of the spherical harmonics, we thus have
\begin{equation}
  a^\text{eq}_{\ell m} = s_{\ell m}
  \label{eq:alm-eq}
\end{equation}
as the result for the first order equilibrium drop shape.  Note that this
implies $a^\text{eq}_{00}=0$.  This will provide a useful point of comparison
for the nonequilibrium shapes we obtain later.

\subsection{Gibbs-Thomson relation for the Wulff shape}

We observe that substituting the equilibrium result $a_{\ell m}^\text{eq}$ into
the Gibbs-Thomson relation (\ref{eq:mu-gt-first-order}) causes the
angular-dependent first order terms to vanish.  This is a reflection
of a stronger statement that can be made without resort to
perturbation theory that the Gibbs-Thomson relation for the
equilibrium drop shape has a uniform value over the surface of the
drop.

The equilibrium shape will minimize the free energy while maintaining
a fixed volume.    The volume of our drop is given by
\begin{equation}
  V[R(\theta,\phi)] = \int d\Omega \frac{R^3}{3}
\end{equation}
while the free energy is given by (\ref{eq:free-energy}).  We minimize
the free energy subject to the constraint of a constant volume by
means of a Lagrange multiplier \cite{Boas2006}.  Introduce
\begin{equation}
  G \equiv F - \lambda V = \int d\Omega\, g(R,R_\theta,R_\phi;\theta,\phi),
\end{equation}
which means $g = f - \lambda R^3/3$, and then minimize $G$ with respect
to a variation $R\to R+\delta R$.  This becomes
\begin{equation}
0=\frac{\partial g}{\partial R}
    - \frac{\partial}{\partial\theta}\frac{\partial g}{\partial R_\theta}
    - \cot\theta\frac{\partial g}{\partial R_\theta} - \frac{\partial}
    {\partial\phi} \frac{\partial g}{\partial R_\phi} =
    \mu \Delta\rho R^2
    - \lambda R^2,
\end{equation}
where we have made use of (\ref{eq:anisotropic-gt}).  The chemical
potential is then a constant $\mu = \lambda/\Delta\rho$ as claimed.

\section{Interface Dynamics}

\label{sec:interface-dynamics}

In this section we determine the interface dynamics to first order in
$\delta$ by solving the Dirichlet problem posed in
Sec.~\ref{subsec:coarsening-phenomenology} for the chemical potential
inside and outside the drop and then relating the discontinuity in the
gradient of the chemical potential to the interface velocity.

In the isotropic LS theory, the time-dependent supersaturation at
infinity provides a nonzero $\mu_\infty$, which was used to define
the critical radius $R_c$.  We choose to preserve the isotropic result
for $R_c$, given by (\ref{eq:Rc}), but we must allow for the supersaturation
at infinity to be modified by the anisotropy.  Thus we take
\begin{equation}
  \mu_\infty = \frac{2\sigma_0}{\Delta\rho R_c}[1 + \delta \nu_1
    +\delta^2\nu_2 +\dots]
  \label{eq:mu-infinity-expansion}
\end{equation}
where the $\nu_i$ are dimensionless constants (i.e., no dependence on
the drop size parameter $R_0$).  We will determine the values of the
$\nu_i$ in Sec.~\ref{sec:drop-size-distribution} by the constraint
that the total volume fraction is $\epsilon$.

We express the interior and exterior solutions, which
both obey Laplace's equation, as expansions
in spherical harmonics \cite{Jackson1999}
\begin{equation}
  \mu^\text{in}({\bf r}) =  \sum_{\ell,m} m_{\ell m}^\text{in} r^\ell
  Y_\ell^m(\theta,\phi)
\end{equation}
and
\begin{equation}
  \mu^\text{out}({\bf r}) = \mu_\infty 
  + \sum_{\ell,m} m_{\ell m}^\text{out} r^{-(\ell+1)}
  Y_\ell^m(\theta,\phi).
\end{equation}
Note that $\mu_{00}^\text{in}$, $\mu_{00}^\text{out}$, and
$\mu_\infty$ all contain $O(\delta^0)$ terms, whereas the leading terms
for $\mu_{\ell m}^\text{in}$ and $\mu_{\ell m}^\text{out}$ with
$\ell>0$ are $O(\delta)$.  We match these expansions to the
Gibbs-Thomson result Eq.~(\ref{eq:mu-gt-first-order}) at the boundary
$r=R(\theta,\phi)$ to obtain
\begin{align}
  \mu^\text{in}({\bf r}) =& \frac{2\sigma_0}{\Delta\rho R_0}\biggl[
    1 + \delta\biggl\{\sum_{\ell m}\beta_\ell
    (a_{\ell m}-s_{\ell m})\biggl(\frac{r}{R_0}\biggr)^\ell Y_\ell^m
    \biggr\}\biggr]\nonumber\\
   &+ O(\delta^2).
\end{align}
and
\begin{align}
  \mu^\text{out}&({\bf r}) = 
  \mu_\infty
  + \frac{2\sigma_0}{\Delta\rho}\frac{1}{r}
   \biggl[ 1 - x + \delta\biggl\{-x\nu_1 \nonumber\\
  &+ \sum_{\ell m}\biggl(\beta_\ell
     (a_{\ell m}-s_{\ell m})+(1-x)a_{\ell m}\biggr)\biggl(\frac{R_0}{r}\biggr)^\ell
     Y_\ell^m
    \biggr\}\biggr]\nonumber\\
   &+ O(\delta^2),
\end{align}
where $x=R_0/R_c$, $\beta_\ell$ is defined in Eq.~(\ref{eq:beta}), and
we have suppressed the arguments of $Y_{\ell m}(\theta,\phi)$ for
brevity.

Now we turn to Eq.~(\ref{eq:velocity}) to obtain the interface
normal velocity.  First, we note that
\begin{equation}
  \hat{\bf n}\cdot\nabla = \frac{\partial}{\partial r} - \frac{R_\theta}{R}
  \frac{\partial}{\partial\theta} - \frac{R_\phi}{R\sin^2\theta}
  \frac{\partial}{\partial\phi},
\end{equation}
where $R$ is given by (\ref{eq:R-expansion}).  Note that $R_\theta$
and $R_\phi$ are of order $\delta$ and that the $\theta$ and
$\phi$ derivatives acting on the chemical potential solutions will
provide additional factors of order $\delta$.  Thus,
\begin{equation}
  v_n = \frac{M}{\Delta\rho}\biggl(\frac{\partial\mu^\text{out}}{\partial r}
   -\frac{\partial\mu^\text{in}}{\partial r}
  \biggr)\bigg|_{r=R(\theta,\phi)}
  + O(\delta^2).
  \label{eq:velocity-expansion}
\end{equation}
We also note that the radial velocity $v_r$ is related to the normal
velocity $v_n$ via
\begin{equation}
  v_r = v_n / (\hat{\bf n}\cdot \hat{\bf r}) = v_n + O(\delta^2).
\end{equation}
Plugging the chemical potential solutions into (\ref{eq:velocity-expansion})
results in
\begin{align}
  v_r =& \frac{\alpha}{R_0}\biggl(\frac{1}{R_c}-\frac{1}{R_0}\biggr)
  + \delta \frac{\alpha}{R_0^2}\biggl[ \nu_1 x  \nonumber\\
    &- \sum_{\ell,m}\biggl\{ \gamma_\ell(a_{\ell m}-s_{\ell m})+(\ell-1)
    (1-x) a_{\ell m}\biggr\} Y_{\ell}^m\biggr]
  \label{eq:vr-expansion}
\end{align}
where $\alpha$ is defined just after Eq.~(\ref{eq:ls-drop-dynamics})
and
\begin{equation}
  \gamma_\ell = (2\ell+1)\beta_\ell = (2\ell+1)[\ell(\ell+1)-2]/2.
\label{eq:gamma-def}
\end{equation}

\section{Drop Center Motion}

\label{sec:drop-center-motion}

To this point we have left ambiguous how we determine the
origin for our drop parametrization $R(\theta,\phi)$.
A simple and straightforward choice
is to impose that all $\ell=1$ components of the drop shape function
Eq.~(\ref{eq:R-expansion})
are identically zero.  I.e., $\rho_{1,m} \equiv \delta a_{1,m} + \delta^2 b_{1,m} +
\dots = 0$ to all orders
in perturbation theory  for $m=-1$, $0$, $1$.
Unsurprisingly,
we obtain three conditions to specify the origin in three dimensions.

To see that this is always possible and represents a unique choice
for the origin, consider starting with some arbitrary choice for the origin
and then define the location of the drop center ${\bf r}_c$ via
\begin{equation}
  {\bf r}_c \equiv \int \frac{d\Omega}{4\pi} R(\theta,\phi)\, \hat{\bf r}
\end{equation}
Now we shift coordinates to a drop parametrization given
by ${\bf R}' = {\bf R}-
{\bf r}_c$.  In these new coordinates, which we are always free to choose,
the drop center will be at the origin and our drop shape will
satisfy the constraint $\int d\Omega\, {\bf R} = 0$, which implies
$\rho_{1,m}=0$.

While we can impose that all $\ell=1$ terms in the drop shape are
identically zero from here on, this requires us to allow for the
possibility that the drop center may move.  In fact, for certain types
of anisotropy, in particular cases where the surface tension expressed
in terms of spherical harmonics has non-zero modes $\ell_1$ and
$\ell_2$ such that $|\ell_1-\ell_2|=1$, it can happen that no point in
space remains in the interior of the drop during its full evolution.
Thus to determine how the drop shape $R(\theta,\phi,t)$ evolves in
time, it becomes necessary to consider the combined effects of both
interface motion and drop center motion. We next develop the equation
of motion for $R$ that accounts for both factors.

Consider at time $t$ the drop center to have velocity ${\bf v}_c$,
which we resolve into spherical coordinates as ${\bf v}_c =
v_{c,r}\hat{\bf r} + v_{c,\theta}\hat{\bm\theta} +
v_{c,\phi}\hat{\bm\phi}$ with the origin located at the instantaneous
position of the drop center.  At some infinitesimally later time
$t+\delta t$ the drop center has moved to the location ${\bf
  v}_c\delta t$, and our goal is to determine how the function
$R(\theta,\phi,t)$ defining the shape and size of the drop has evolved
due to the combined effects of interface motion and drop center
motion.

\begin{figure}
  \includegraphics[width=2.8in]{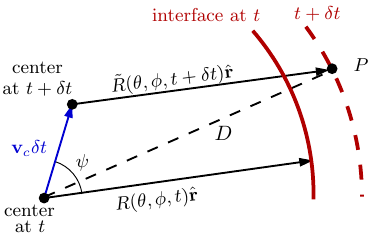}
  \caption{Update in the drop parametrization $R(\theta,\phi,t)$ due
    to combined effects of interface motion and drop center motion.
    This diagram is drawn in the plane defined by ${\bf v}_c$ and
    $\hat{\bf r}$.}
  \label{fig:drop-motion}
\end{figure}

We fix $\theta$ and $\phi$ and focus on the quantity $\tilde
R(\theta,\phi,t+\delta)$, indicated in Fig.~\ref{fig:drop-motion},
which points in the original $\hat{\bf r}$ direction from the new drop
center location at time $t+\delta t$ to the updated interface
location, labelled as point $P$.  Once we obtain an expression for this
quantity to order $\delta t$, we can use it to solve for the time
evolution of our drop via
\begin{equation}
  \dot R(\theta,\phi,t) = \frac{\tilde R(\theta,\phi,t+\delta t)-
    R(\theta,\phi,t)}{\delta t},
  \label{eq:drop-evolution-def}
\end{equation}
for infinitesimal $\delta t$.

We define $\delta\theta$ and $\delta\phi$
via the direction  indicated by the
dashed line in Fig.~\ref{fig:drop-motion} from the
original drop
center to the point $P$, which we take to be
$\hat{\bf r}(\theta+\delta\theta,\phi+\delta\phi)$. We presume both
$\delta\theta$ and $\delta\phi$ to be of order
$\delta t$.
The quantity $D$ in Fig.~\ref{fig:drop-motion} can then
be written as
\begin{equation}
  D = R + R_\theta\delta\theta +R_\phi\delta\phi + v_r\delta t + O(\delta t^2)
\end{equation}
where all quantities are evaluated at $\theta$, $\phi$, and $t$.  We then
use the law of cosines to obtain $\tilde R$ in terms of $D$ and $v_c\delta t$,
giving
\begin{equation}
  \tilde R = R + R_\theta \delta\theta + R_\phi\delta\phi + v_r\delta t
  - v_c \cos\psi \delta t
  \label{eq:law-of-cosines}
\end{equation}
where again we have kept only leading order terms in $\delta t$, and
have introduced $\psi$ as the angle between ${\bf v}_c$ and $\hat{\bf r}$.

Finally, we note that $\delta\theta$ and $\delta\phi$ are affected by
both the motion of the drop center and the motion of the interface, but the
latter is a higher order correction.  Thus to leading order,
\begin{eqnarray}
  R\delta\theta  &= v_{c,\theta}\delta t\\
  R\sin\theta\delta\phi &= v_{c,\phi}\delta t.
\end{eqnarray}
Inserting this into Eqs.~(\ref{eq:drop-evolution-def}) and
(\ref{eq:law-of-cosines}) and using $v_c\cos\psi = v_{c,r}$ gives
the equation governing drop evolution with a moving drop center:
\begin{equation}
  \dot R = v_r + \frac{R_\theta}{R}v_{c,\theta} + \frac{R_\phi}{R\sin\theta}
  v_{c,\phi} - v_{c,r}.
\end{equation}
This is the main result of this section, which describes how the
equation relating $\dot R$ and $v_r$ is modified by the drop motion
${\bf v}_c$.

To use this, we want to impose that $\dot R$ expressed in
spherical harmonics has no $\ell=1$ term.  This will ensure that our
drop shape will preserve the property $\rho_{1,m}=0$.  When we
solve for $v_r$ perturbatively, then any non-zero $\ell =1$ term can
be canceled by adjusting $v_{c,\theta}$, $v_{c,\phi}$, and $v_{c,r}$
as needed.


\section{Scaling Solution for the Drop Shape}

\label{sec:scaling-solution}

Starting from the expansion for $R(\theta,\phi)$ in Eq.~(\ref{eq:R-expansion})
we develop the equations of evolution for the coefficients
$a_{\ell m}$, $b_{\ell m}$, etc.  We are looking for scaling solutions, where
\begin{equation}
  a_{\ell m}(R_0,t) = a_{\ell m}(x),
\end{equation}
where as before, $x=R_0/R_c(t)$.
Taking the time derivative we obtain
\begin{equation}
  \dot R = \dot R_0 + \delta \sum_{\ell m} \biggl(
  \dot R_0 a_{\ell m}(x) + R_0 a_{\ell m}'(x) \dot x\biggr) Y_\ell^m
  + O(\delta^2).
\label{eq:Rdot-first-order}
\end{equation}
Since drop motion only affects $\dot R$ at second order, we have $\dot
R = v_r + O(\delta^2)$, and so we equate Eqs.~(\ref{eq:Rdot-first-order})
and (\ref{eq:vr-expansion}) to obtain
\begin{equation}
  \dot R_0 a_{00} + R_0 a'_{00}\dot x = \frac{\alpha}{R_0^2}
  \Bigl( \nu_1 x + (2-x) a_{00}\Bigr)
\end{equation}
for the $\ell=0$ term and
\begin{equation}
  \dot R_0 a_{\ell m} + R_0 a_{\ell m}' \dot x =
  -\frac{\alpha}{R_0^2}\Bigl[ \gamma_\ell(a_{\ell m}-s_{\ell m})
    + (\ell-1)(1-x) a_{\ell m}\Bigr]
\end{equation}
for $\ell>0$.
In these expressions $\dot R_0$ is given by the isotropic Lifshitz-Slyozov
radial velocity in Eq.~(\ref{eq:ls-drop-dynamics}), $R_c(t)$ is given by
Eq.~(\ref{eq:Rc}), and
\begin{equation}
  \dot x = \frac{\dot R_0}{R_c} - x\frac{\dot R_c}{R_c}
  =  -\frac{(x+3)(\frac{3}{2}-x)^2}{3 x^2 t}.
\end{equation}

With these substitutions the equation for $a_{00}(x)$ becomes
\begin{equation}
  x (x+3)(3-2x)^2 a_{00}' + 27(3-2x) a_{00} = -27 x\,\nu_1.
\end{equation}
If we require on physical grounds that $a_{00}$ and $a_{00}'$ are
finite for $0\leq x\leq 3/2$, then the limit of $x\to 3/2$ in the
equation above implies that $\nu_1 = 0$.  Thus the leading correction
to $\mu_\infty$ in Eq.~(\ref{eq:mu-infinity-expansion}) must
come in at second order in anisotropy strength.

The resulting homogeneous equation for $a_{00}$ has no solution that
is finite on the interval $0\leq x\leq 3/2$ except for $a_{00}=0$, so
we conclude there is no correction to the drop size to first order
in the anisotropy strength.

For $\ell>0$ the equation for $a_{\ell m}$ is given by
\begin{equation}
  \frac{x(x+3)(3-2x)^2}{27} a_{\ell m}' +
       [2-\gamma_\ell - \ell + (\ell -2) x] a_{\ell m} =
       -\gamma_\ell s_{\ell m}.
\end{equation}
This first order equation can be solved by the method of integration
factors to give
\begin{equation}
  a_{\ell m}(x) = e^{-F(x)}\biggl(a_{\ell m}(x_0) e^{F(x_0)}
  + \int_{x_0}^x e^{F(y)} g(y)\,dy\biggr)
\end{equation}
where the integration factor is
\begin{equation}
  e^{-F(x)} = \frac{x^{\gamma_\ell+\ell-2}
  \exp\Bigl(
  \frac{\gamma_\ell+1-\ell/2}{{\textstyle\frac{3}{2}}-x}\Bigr)}
  {(x+3)^{(\gamma_\ell+4\ell-8)/9}
  ({\textstyle\frac{3}{2}}-x)^{(8\gamma_\ell+5\ell-10)/9}} ,
\end{equation}
and
\begin{equation}
  g(y) = -\frac{27\gamma_\ell s_{\ell m}}{x(x+3)(3-2x)^2}.
\end{equation}
Since $e^{-F(x)}$ diverges as $x\to 3/2$, we choose $x_0=3/2$ to eliminate
the boundary term, giving
\begin{equation}
  a_{\ell m}(x) = - e^{-F(x)}\int_x^{3/2} e^{F(y)} g(y) \, dy
  \label{eq:alm-solution}
\end{equation}
where $F(x)$ and $g(y)$ are given above.

Eq.~(\ref{eq:alm-solution}) is the primary result of this paper.
It demonstrates that the $a_{\ell m}$ drop shape
coefficients are functions only of $x$, thereby exhibiting scaling and
providing a one-parameter family of
nonequilibrium drop shapes that are unequal to the Wulff shape.

\begin{figure}
  \includegraphics[width=3.2in]{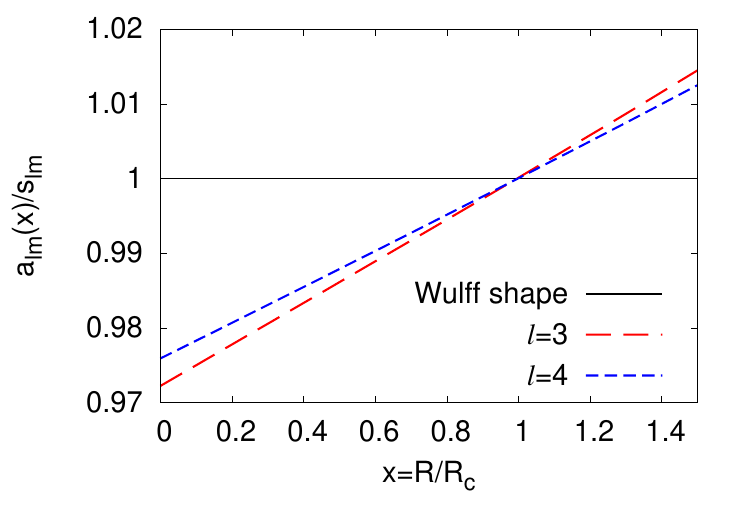}
  \caption{The scaling solution for $a_{\ell m}(x)$ as compared to the
    Wulff shape.  These represent a one-parameter family of nonequilibrium
  shapes.}
  \label{fig:alm-solution}
\end{figure}

A few observations are in order.  This solution is finite and analytic
on the interval $0\leq x\leq 3/2$, with the limiting values
\begin{equation}
  a_{\ell m}(0) = \frac{\gamma_\ell}{\gamma_\ell + \ell -2} s_{\ell m}
\end{equation}
and
\begin{equation}
  a_{\ell m}(3/2) = \frac{\gamma_\ell}{\gamma_\ell-  \ell/2 +1} s_{\ell m}.
\end{equation}
For $\ell=2$ the solution reduces to $a_{2m} = s_{2m}$.  This follows
directly from
the observation that, for $\ell=2$, $g(y) = s_{2m} dF(y)/dy$.  A
numerical solution is  plotted in  Fig.~\ref{fig:alm-solution} for
$\ell=3$ and $4$.
The plot shows that the $a_{\ell m}$ coefficients differ at most
only a few percent from the equilibrium values $a_{\ell m}^\text{eq}=s_{\ell m}$.
The smaller drops are more spherical than the Wulff shape, while the larger
drops have stronger anisotropy than the Wulff shape.

\section{Drop Size Distribution}

\label{sec:drop-size-distribution}

In this section we will address the question of whether the drop size
distribution is modified from the isotropic result
(\ref{eq:ls-distribution}).  We will also clarify how to determine the
coefficients $\nu_i$ in the expansion of $\mu_\infty$,
Eq.~(\ref{eq:mu-infinity-expansion}).  For both tasks we will need to
evaluate the volume
\begin{equation}
  V[R(\theta,\phi)] = \frac{1}{3}\int d\Omega \, R(\theta,\phi)^3.
\end{equation}
given the drop shape expansion Eq.~(\ref{eq:R-expansion}).
We make use of the spherical harmonics identity
\begin{equation}
  \int d\Omega\, Y_\ell^m(\Omega) Y_{\ell'}^{m'}(\Omega) = (-1)^m
  \delta_{\ell,\ell'} \delta_{m,-m'},
\end{equation}
which follows from the conjugation relation $Y_\ell^{m*}(\Omega)
= (-1)^mY_\ell^m(\Omega)$ and the orthogonality condition, regardless
of phase convention.  The volume then evaluates to
\begin{equation}
  V = \frac{4\pi R_0^3}{3}\biggl[ 1 + \delta \frac{3a_{00}}{\sqrt{4\pi}}
    + \delta^2\biggl(\frac{3b_{00}}{\sqrt{4\pi}} +
    3\sum_{\ell m}\frac{(-1)^m a_{\ell m}a_{\ell,-m}}{4\pi}\biggr)+\dots\biggr]
  \label{eq:V-expansion}
\end{equation}
Note that this volume is a function of $R_0$ and $R_c$, since the
$a_{\ell m}$, $b_{\ell m}$, \dots are functions of $x=R_0/R_c$.

The coefficients $\nu_i$ that appear in
Eq.~(\ref{eq:mu-infinity-expansion}), the expansion for the chemical
potential at infinity, are determined by the volume fraction
normalization constraint obeyed by the minority phase, namely
\begin{equation}
  \int\frac{dR_0}{R_c^4} f(R_0/R_c) V(R_0,R_c) = \epsilon.
\end{equation}
Recall that we have constructed our perturbation theory such that
$f(x)$ is the isotropic LS distribution.  We insert the expansion
(\ref{eq:V-expansion}) into this constraint and evaluate order by
order.  The zeroth order constraint just reproduces the isotropic
normalization condition,
\begin{equation}
  \int dx \, f(x) \frac{4\pi}{3} x^3 = \epsilon.
\end{equation}
Thus all higher order terms must contribute zero, i.e., to first order
we have
\begin{equation}
  \int dx \, f(x) \frac{a_{00}(x)}{4\pi} x^3 = 0.
\end{equation}
In principle, this constraint would determine the value of $\nu_1$, but we
already found that the only finite solution possible for $a_{00}$ was
for $\nu_1$ and $a_{00}$ both to vanish.

To second order, the constraint becomes
\begin{equation}
  \int dx\, f(x) \biggl(\frac{b_{00}}{\sqrt{4\pi}}
   + \sum_{\ell>0,m}\frac{a_{\ell m}a_{\ell,-m}}{4\pi}\biggr) x^3 = 0.
\end{equation}
We have obtained non-zero solutions for the $a_{\ell m}(x)$.  While we
have not solved for $b_{00}(x)$, it is clear that it will contain a term
that is linear in $\nu_2$, thus the integral above serves to determine
the value of $\nu_2$.  This program can be carried out order by order
in the anisotropy strength.
Physically, this makes sense.  The value of the supersaturation at infinity,
and therefore the chemical potential at infinity, is directly linked to the
volume fraction of the minority phase.

Now we proceed to address the drop size distribution for the anisotropic
case.  It is necessary to define
a size for the anisotropic drop analogous to the radius of a sphere.
There are many ways to do this, such as using the volume of the
drop, the surface area, or the $n$th moment of $R(\theta,\phi)$.  We will
define the ``volume radius'' to be
$R_V = (3V/4\pi)^{1/3}$, which results in
\begin{equation}
  R_V = R_0\biggl[ 1 
    + \delta^2\biggl( \frac{b_{00}}{\sqrt{4\pi}} +
    \sum_{\ell>0,m} \frac{(-1)^m a_{\ell m} a_{\ell,-m}}{4\pi}\biggr) + \dots \biggr],
  \label{eq:Rv}
\end{equation}
where we have set $a_{00}=0$.  We define the scaled drop size
\begin{equation}
  z \equiv R_V/R_c(t)
\end{equation}
and observe that Eq.~(\ref{eq:Rv}) provides $z$ as a function of $x=R_0/R_c$.
Since we know the distribution for $x$ is the isotropic LS distribution,
we can obtain the $z$ distribution from $f(x)\,dx = f_V(z)\,dz$, which gives
\begin{equation}
  f_V(z) = f\Bigl(x(z)\Bigr) \biggl(\frac{dz}{dx}\biggr)^{-1}.
\end{equation}
Thus, unless $z=x$, the drop size distribution will be modified from the
isotropic case.  To obtain $z=x$ it would be necessary for the second
order term in Eq.~(\ref{eq:Rv}) to be identically zero, which appears
to be unlikely.

\section{Summary}

\label{sec:summary}


This work revisited the landmark Lifshitz-Slyozov theory to
investigate the impact of an anisotropic surface tension on late-stage
coarsening dynamics in the dilute limit.  Our findings, derived
through a perturbative treatment for weak anisotropy, contradict the
earlier expectations of Lifshitz and Slyozov that anisotropy in the
surface tension would lead to uniform anisotropic drop shape and that the
distribution function would remain unaffected.

To demonstrate this, we utilized an expansion of the anisotropic
surface tension $\sigma(\mathbf{n})$ and the drop shape $R(\theta,
\phi)$ in spherical harmonics, developed the anisotropic Gibbs-Thomson
relation, and derived the resulting interface dynamics, accounting for
drop center motion.

Our primary results demonstrate a significant breakdown of
morphological universality.  We found that surface tension anisotropy
results in a one-parameter family of nonequilibrium drop shapes that
are dependent on the scaled drop size $x = R_0/R_c(t)$.  We provided
an explicit solution for the scaling form of the shape coefficients
$a_{\ell m}$ to first order in anisotropy strength. These scaling
shapes are unequal to the equilibrium Wulff shape, with smaller drops
($x \to 0$) more spherical than the corresponding Wulff shape, while
larger drops ($x \to 3/2$) exhibit stronger anisotropy than the Wulff
shape.  Despite the shape modification, the characteristic $t^{1/3}$
drop growth law remains unchanged.

We also confirmed that the drop size distribution is modified by
anisotropy.  Using the volume radius $R_V = (3V/4\pi)^{1/3}$ to define
the anisotropic drop size, with scaled drop size $z = R_V/R_c(t)$, the
resulting distribution $f_V(z)$ will be modified from the isotropic LS
distribution $f(x)$, barring an improbable cancellation of terms.

Future work could be to extend the calculation to second order, at
least for the isotropic $b_{00}$ term, to provide an explicit
expression for the anisotropic drop size distribution.  This would be a
significant undertaking, beyond the scope of the present work.  Our
anisotropy parametrization and perturbative framework could
prove useful for other situations that give rise to weakly
anisotropic drops, such as a weak applied shear or a gravitational
field.  Finally, the question of how an anisotropic surface tension
affects the domain morphology should be further explored outside of the
dilute limit, which will require numerical methods.

\appendix

\section{Anisotropic Gibbs-Thomson Evaluation}

\label{sec:appendix}

Here we outline the evaluation of Eq.~(\ref{eq:anisotropic-gt}).  To
compute the derivatives of $f(R,R_\theta,R_\phi; \theta,\phi)$ it is
helpful to introduce
\begin{equation}
  \Delta \equiv R^2 + R_\theta^2 + R_\phi^2\csc^2\theta = \frac{R^2}{
    (\hat{\bf n}\cdot\hat{\bf r})^2}
\end{equation}
which gives $f=R\sigma(\theta_n,\phi_n)/\Delta^{1/2}$.
Now we can evaluate the partial derivatives
\begin{equation}
  \frac{\partial f}{\partial R} = \frac{\sigma}{\Delta^{1/2}}
  + \frac{R}{\Delta^{1/2}}\biggl[
    \sigma^{(1,0)}\frac{\partial\theta_n}{\partial R} +
    \sigma^{(0,1)}\frac{\partial\phi_n}{\partial R} \biggr]
  -\frac{R^2\sigma}{\Delta^{3/2}}
\end{equation}
\begin{equation}
  \frac{\partial f}{\partial R_\theta} =
   \frac{R}{\Delta^{1/2}}\biggl[
    \sigma^{(1,0)}\frac{\partial\theta_n}{\partial R_\theta} +
    \sigma^{(0,1)}\frac{\partial\phi_n}{\partial R_\theta} \biggr]
   -\frac{R\sigma R_\theta}{\Delta^{3/2}}
\end{equation}
\begin{equation}
  \frac{\partial f}{\partial R_\phi} =
   \frac{R}{\Delta^{1/2}}\biggl[
    \sigma^{(1,0)}\frac{\partial\theta_n}{\partial R_\phi} +
    \sigma^{(0,1)}\frac{\partial\phi_n}{\partial R_\phi} \biggr]
  -\frac{R\sigma R_\phi}{\sin^2\theta\Delta^{3/2}},
\end{equation}
where
\begin{equation}
  \sigma^{(i,j)} = \biggl(\frac{\partial}{\partial\theta_n}\biggr)^i
  \biggl(\frac{\partial}{\partial\phi_n}\biggr)^j
  \sigma(\theta_n,\phi_n).
\end{equation}
The normal angles $\theta_n$ and $\phi_n$ are given by
Eqs.~(\ref{eq:thetan}) and (\ref{eq:phin}), and have the derivatives
\begin{equation}
  \frac{\partial\theta_n}{\partial R} = \frac{RR_\theta\sin\theta
    - R_\theta^2\cos\theta - R_\phi^2\cos\theta\csc^2\theta}{
    \Delta\bigl[\Delta - (R\cos\theta+R_\theta\sin\theta)^2\bigr]^{1/2}}
\end{equation}
\begin{equation}
  \frac{\partial\theta_n}{\partial R_\theta} = \frac{RR_\theta\cos\theta
    - R_\phi^2\csc\theta - R^2\sin\theta}{
    \Delta\bigl[\Delta - (R\cos\theta+R_\theta\sin\theta)^2\bigr]^{1/2}}
\end{equation}
\begin{equation}
  \frac{\partial\theta_n}{\partial R_\phi} = \frac{R_\phi R_\theta\csc\theta
    + R R_\phi \cot\theta\csc\theta}{
    \Delta\bigl[\Delta - (R\cos\theta+R_\theta\sin\theta)^2\bigr]^{1/2}}
\end{equation}
and
\begin{equation}
  \frac{\partial\phi_n}{\partial R} = \frac{R_\phi}
       {R_\phi^2\csc^2\theta + (R_\theta\cos\theta - R\sin\theta)^2}
\end{equation}
\begin{equation}
  \frac{\partial\phi_n}{\partial R_\theta} = -\frac{R_\phi\cot\theta}
       {R_\phi^2\csc^2\theta + (R_\theta\cos\theta - R\sin\theta)^2}
\end{equation}
\begin{equation}
  \frac{\partial\phi_n}{\partial R_\phi} = \frac{R-R_\theta\cot\theta}
       {R_\phi^2\csc^2\theta + (R_\theta\cos\theta - R\sin\theta)^2}
\end{equation}
When evaluating the $\partial/\partial\theta$ and $\partial/\partial\phi$
derivatives in Eq.~(\ref{eq:anisotropic-gt}) it is important to note
that they act on the implicit angle dependence of $R$, $R_\theta$, and $R_\phi$,
as well as the explicit angles in the expressions above.

After taking these derivatives, the next step in the evaluation is to
express $\sigma^{(i,j)}(\theta_n,\phi_n)$ in terms of $\theta$ and
$\phi$.  For this purpose we need the perturbative expansions
\begin{equation}
  \theta_n = \theta - \delta \frac{R_\theta}{R}
  + \delta^2 \frac{R_\phi^2\cos\theta}{2R^2\sin^3\theta} + O(\delta^3)
\end{equation}
\begin{equation}
  \phi_n = \phi - \delta \frac{R_\phi}{R\sin^2\theta}
  - \delta^2 \frac{R_\theta R_\phi \cos\theta}{R^2\sin^3\theta} + O(\delta^3),
\end{equation}
which are obtained from Eqs.~(\ref{eq:thetan}) and (\ref{eq:phin}).
The resulting expressions were handled
within a Mathematica notebook and are too unwieldy to reproduce here.

We note two features of the resulting expressions.  The first is that if
we take an isotropic surface tension $\sigma=\sigma_0$ (equivalently, all
$s_{\ell m}=0$), the result for $\mu$ becomes
\begin{equation}
  \mu = \frac{\sigma_0}{\Delta\rho}\biggl(\frac{A\sin^3\theta
    + B \sin^2\theta\cos\theta + C \sin\theta + D \cos\theta}
      {R\Delta^{3/2}\sin^3\theta}\biggr)
\end{equation}
with
\begin{align}
  A =& 3RR_\theta^2 + 2 R^3 - R^2R_{\theta\theta} \\
  B =& -R^2R_\theta - R_\theta^3 \\
  C =& 3RR_\phi^2 - R_{\theta\theta}R_\phi^2 + 2 R_\theta R_\phi R_{\theta\phi}
  - R^2R_\phi\nonumber\\
  &- R_\theta^2R_\phi \\
  D =& -2R_\theta R_\phi^2
\end{align}
This exactly matches the expression derived for the mean curvature of
an anisotropic drop in \cite{Mazharimousavi2017}, as we would expect
based on Eq.~(\ref{eq:anisotropic-gt-classic}).

The second feature is that when we expand our expression in powers of
$\delta$, we obtain
\begin{equation}
  \mu = \frac{\sigma_0}{\Delta\rho R_0}\biggl[2
    - \delta \sum_{\ell m} (a_{\ell m}-s_{\ell m})
    (2+\nabla^2_{S^2})Y_\ell^m(\theta,\phi) + O(\delta^2) \biggr],
\end{equation}
where $\nabla^2_{S^2} = \partial_\theta^2 + \cot\theta\partial_\theta
+ \csc^2\theta\partial\phi^2$ is the Laplace-Beltrami operator.
Spherical harmonics are eigenfunctions of this operator,
$\nabla^2_{S^2} Y_\ell^m = -\ell(\ell+1) Y_\ell^m$, which provides
our result Eq.~(\ref{eq:anisotropic-gt}).

\begin{acknowledgments}
  BPV-L would like to acknowledge many fruitful discussions with
  Andrew Rutenberg.
\end{acknowledgments}


\begin{thebibliography}{31}%
\makeatletter
\providecommand \@ifxundefined [1]{%
 \@ifx{#1\undefined}
}%
\providecommand \@ifnum [1]{%
 \ifnum #1\expandafter \@firstoftwo
 \else \expandafter \@secondoftwo
 \fi
}%
\providecommand \@ifx [1]{%
 \ifx #1\expandafter \@firstoftwo
 \else \expandafter \@secondoftwo
 \fi
}%
\providecommand \natexlab [1]{#1}%
\providecommand \enquote  [1]{``#1''}%
\providecommand \bibnamefont  [1]{#1}%
\providecommand \bibfnamefont [1]{#1}%
\providecommand \citenamefont [1]{#1}%
\providecommand \href@noop [0]{\@secondoftwo}%
\providecommand \href [0]{\begingroup \@sanitize@url \@href}%
\providecommand \@href[1]{\@@startlink{#1}\@@href}%
\providecommand \@@href[1]{\endgroup#1\@@endlink}%
\providecommand \@sanitize@url [0]{\catcode `\\12\catcode `\$12\catcode
  `\&12\catcode `\#12\catcode `\^12\catcode `\_12\catcode `\%12\relax}%
\providecommand \@@startlink[1]{}%
\providecommand \@@endlink[0]{}%
\providecommand \url  [0]{\begingroup\@sanitize@url \@url }%
\providecommand \@url [1]{\endgroup\@href {#1}{\urlprefix }}%
\providecommand \urlprefix  [0]{URL }%
\providecommand \Eprint [0]{\href }%
\providecommand \doibase [0]{https://doi.org/}%
\providecommand \selectlanguage [0]{\@gobble}%
\providecommand \bibinfo  [0]{\@secondoftwo}%
\providecommand \bibfield  [0]{\@secondoftwo}%
\providecommand \translation [1]{[#1]}%
\providecommand \BibitemOpen [0]{}%
\providecommand \bibitemStop [0]{}%
\providecommand \bibitemNoStop [0]{.\EOS\space}%
\providecommand \EOS [0]{\spacefactor3000\relax}%
\providecommand \BibitemShut  [1]{\csname bibitem#1\endcsname}%
\let\auto@bib@innerbib\@empty
\bibitem [{\citenamefont {Gunton}\ \emph {et~al.}(1983)\citenamefont {Gunton},
  \citenamefont {San~Miguel},\ and\ \citenamefont {Sahni}}]{Gunton1983}%
  \BibitemOpen
  \bibfield  {author} {\bibinfo {author} {\bibfnamefont {J.}~\bibnamefont
  {Gunton}}, \bibinfo {author} {\bibfnamefont {M.}~\bibnamefont {San~Miguel}},\
  and\ \bibinfo {author} {\bibfnamefont {P.}~\bibnamefont {Sahni}},\ }\bibinfo
  {title} {The dynamics of first order phase transitions},\ in\ \href@noop {}
  {\emph {\bibinfo {booktitle} {Phase Transitions and Critical Phenomena}}},\
  \bibinfo {editor} {edited by\ \bibinfo {editor} {\bibfnamefont
  {C.}~\bibnamefont {Domb}}\ and\ \bibinfo {editor} {\bibfnamefont
  {J.}~\bibnamefont {Lebowitz}}}\ (\bibinfo  {publisher} {Academic},\ \bibinfo
  {address} {New York},\ \bibinfo {year} {1983})\ pp.\ \bibinfo {pages}
  {267--466}\BibitemShut {NoStop}%
\bibitem [{\citenamefont {Bray}(1994)}]{Bray1994}%
  \BibitemOpen
  \bibfield  {author} {\bibinfo {author} {\bibfnamefont {A.~J.}\ \bibnamefont
  {Bray}},\ }\bibfield  {title} {\bibinfo {title} {Theory of phase ordering
  kinetics},\ }\href {https://doi.org/10.1080/00018739400101505} {\bibfield
  {journal} {\bibinfo  {journal} {Adv. Phys.}\ }\textbf {\bibinfo {volume}
  {43}},\ \bibinfo {pages} {357} (\bibinfo {year} {1994})}\BibitemShut
  {NoStop}%
\bibitem [{\citenamefont {Cahn}\ and\ \citenamefont
  {Novick-Cohen}(1994)}]{Cahn1994}%
  \BibitemOpen
  \bibfield  {author} {\bibinfo {author} {\bibfnamefont {J.~W.}\ \bibnamefont
  {Cahn}}\ and\ \bibinfo {author} {\bibfnamefont {A.}~\bibnamefont
  {Novick-Cohen}},\ }\bibfield  {title} {\bibinfo {title} {Evolution equations
  for phase separation and ordering in binary alloys},\ }\href@noop {}
  {\bibfield  {journal} {\bibinfo  {journal} {J. Stat. Phys.}\ }\textbf
  {\bibinfo {volume} {76}},\ \bibinfo {pages} {877} (\bibinfo {year}
  {1994})}\BibitemShut {NoStop}%
\bibitem [{\citenamefont {Orlikowski}\ \emph {et~al.}(1999)\citenamefont
  {Orlikowski}, \citenamefont {Sagui}, \citenamefont {Somoza},\ and\
  \citenamefont {Roland}}]{Orlikowski1999}%
  \BibitemOpen
  \bibfield  {author} {\bibinfo {author} {\bibfnamefont {D.}~\bibnamefont
  {Orlikowski}}, \bibinfo {author} {\bibfnamefont {C.}~\bibnamefont {Sagui}},
  \bibinfo {author} {\bibfnamefont {A.}~\bibnamefont {Somoza}},\ and\ \bibinfo
  {author} {\bibfnamefont {C.}~\bibnamefont {Roland}},\ }\bibfield  {title}
  {\bibinfo {title} {Large-scale simulations of phase separation of elastically
  coherent binary alloy systems},\ }\href
  {https://doi.org/10.1103/PhysRevB.59.8646} {\bibfield  {journal} {\bibinfo
  {journal} {Phys. Rev. B}\ }\textbf {\bibinfo {volume} {59}},\ \bibinfo
  {pages} {8646} (\bibinfo {year} {1999})}\BibitemShut {NoStop}%
\bibitem [{\citenamefont {Stavans}(1993)}]{Stavans1993}%
  \BibitemOpen
  \bibfield  {author} {\bibinfo {author} {\bibfnamefont {J.}~\bibnamefont
  {Stavans}},\ }\bibfield  {title} {\bibinfo {title} {The evolution of cellular
  structures},\ }\href {https://doi.org/10.1088/0034-4885/56/6/002} {\bibfield
  {journal} {\bibinfo  {journal} {Rep. Prog. Phys.}\ }\textbf {\bibinfo
  {volume} {56}},\ \bibinfo {pages} {733} (\bibinfo {year} {1993})}\BibitemShut
  {NoStop}%
\bibitem [{\citenamefont {Hilgenfeldt}\ \emph
  {et~al.}(2001{\natexlab{a}})\citenamefont {Hilgenfeldt}, \citenamefont
  {Kraynik}, \citenamefont {Koehler},\ and\ \citenamefont
  {Stone}}]{Hilgenfeldt2001b}%
  \BibitemOpen
  \bibfield  {author} {\bibinfo {author} {\bibfnamefont {S.}~\bibnamefont
  {Hilgenfeldt}}, \bibinfo {author} {\bibfnamefont {A.~M.}\ \bibnamefont
  {Kraynik}}, \bibinfo {author} {\bibfnamefont {S.~A.}\ \bibnamefont
  {Koehler}},\ and\ \bibinfo {author} {\bibfnamefont {H.~A.}\ \bibnamefont
  {Stone}},\ }\bibfield  {title} {\bibinfo {title} {An accurate von neumann's
  law for three-dimensional foams},\ }\href
  {https://doi.org/10.1103/PhysRevLett.86.2685} {\bibfield  {journal} {\bibinfo
   {journal} {Phys. Rev. Lett.}\ }\textbf {\bibinfo {volume} {86}},\ \bibinfo
  {pages} {2685} (\bibinfo {year} {2001}{\natexlab{a}})}\BibitemShut {NoStop}%
\bibitem [{\citenamefont {Hilgenfeldt}\ \emph
  {et~al.}(2001{\natexlab{b}})\citenamefont {Hilgenfeldt}, \citenamefont
  {Koehler},\ and\ \citenamefont {Stone}}]{Hilgenfeldt2001}%
  \BibitemOpen
  \bibfield  {author} {\bibinfo {author} {\bibfnamefont {S.}~\bibnamefont
  {Hilgenfeldt}}, \bibinfo {author} {\bibfnamefont {S.~A.}\ \bibnamefont
  {Koehler}},\ and\ \bibinfo {author} {\bibfnamefont {H.~A.}\ \bibnamefont
  {Stone}},\ }\bibfield  {title} {\bibinfo {title} {Dynamics of coarsening
  foams: Accelerated and self-limiting drainage},\ }\href
  {https://doi.org/10.1103/PhysRevLett.86.4704} {\bibfield  {journal} {\bibinfo
   {journal} {Phys. Rev. Lett.}\ }\textbf {\bibinfo {volume} {86}},\ \bibinfo
  {pages} {4704} (\bibinfo {year} {2001}{\natexlab{b}})}\BibitemShut {NoStop}%
\bibitem [{\citenamefont {Thomas}\ \emph {et~al.}(2006)\citenamefont {Thomas},
  \citenamefont {de~Almeida},\ and\ \citenamefont {Graner}}]{Thomas2006}%
  \BibitemOpen
  \bibfield  {author} {\bibinfo {author} {\bibfnamefont {G.~L.}\ \bibnamefont
  {Thomas}}, \bibinfo {author} {\bibfnamefont {R.~M.~C.}\ \bibnamefont
  {de~Almeida}},\ and\ \bibinfo {author} {\bibfnamefont {F.}~\bibnamefont
  {Graner}},\ }\bibfield  {title} {\bibinfo {title} {Coarsening of
  three-dimensional grains in crystals, or bubbles in dry foams, tends towards
  a universal, statistically scale-invariant regime},\ }\href
  {https://doi.org/10.1103/PhysRevE.74.021407} {\bibfield  {journal} {\bibinfo
  {journal} {Phys. Rev. E}\ }\textbf {\bibinfo {volume} {74}},\ \bibinfo
  {pages} {021407} (\bibinfo {year} {2006})}\BibitemShut {NoStop}%
\bibitem [{\citenamefont {Lambert}\ \emph {et~al.}(2010)\citenamefont
  {Lambert}, \citenamefont {Mokso}, \citenamefont {Cantat}, \citenamefont
  {Cloetens}, \citenamefont {Glazier}, \citenamefont {Graner},\ and\
  \citenamefont {Delannay}}]{Lambert2010}%
  \BibitemOpen
  \bibfield  {author} {\bibinfo {author} {\bibfnamefont {J.}~\bibnamefont
  {Lambert}}, \bibinfo {author} {\bibfnamefont {R.}~\bibnamefont {Mokso}},
  \bibinfo {author} {\bibfnamefont {I.}~\bibnamefont {Cantat}}, \bibinfo
  {author} {\bibfnamefont {P.}~\bibnamefont {Cloetens}}, \bibinfo {author}
  {\bibfnamefont {J.~A.}\ \bibnamefont {Glazier}}, \bibinfo {author}
  {\bibfnamefont {F.}~\bibnamefont {Graner}},\ and\ \bibinfo {author}
  {\bibfnamefont {R.}~\bibnamefont {Delannay}},\ }\bibfield  {title} {\bibinfo
  {title} {Coarsening foams robustly reach a self-similar growth regime},\
  }\href {https://doi.org/10.1103/PhysRevLett.104.248304} {\bibfield  {journal}
  {\bibinfo  {journal} {Phys. Rev. Lett.}\ }\textbf {\bibinfo {volume} {104}},\
  \bibinfo {pages} {248304} (\bibinfo {year} {2010})}\BibitemShut {NoStop}%
\bibitem [{\citenamefont {Singh}\ and\ \citenamefont
  {Banerjee}(2023)}]{Singh2023}%
  \BibitemOpen
  \bibfield  {author} {\bibinfo {author} {\bibfnamefont {A.~K.}\ \bibnamefont
  {Singh}}\ and\ \bibinfo {author} {\bibfnamefont {V.}~\bibnamefont
  {Banerjee}},\ }\bibfield  {title} {\bibinfo {title} {Phase separation of a
  magnetic fluid: Asymptotic states and nonequilibrium kinetics},\ }\href
  {https://doi.org/10.1103/PhysRevE.108.064604} {\bibfield  {journal} {\bibinfo
   {journal} {Phys. Rev. E}\ }\textbf {\bibinfo {volume} {108}},\ \bibinfo
  {pages} {064604} (\bibinfo {year} {2023})}\BibitemShut {NoStop}%
\bibitem [{\citenamefont {Chakrabarti}\ \emph {et~al.}(1990)\citenamefont
  {Chakrabarti}, \citenamefont {Toral}, \citenamefont {Gunton},\ and\
  \citenamefont {Muthukumar}}]{Chakrabarti1990}%
  \BibitemOpen
  \bibfield  {author} {\bibinfo {author} {\bibfnamefont {A.}~\bibnamefont
  {Chakrabarti}}, \bibinfo {author} {\bibfnamefont {R.}~\bibnamefont {Toral}},
  \bibinfo {author} {\bibfnamefont {J.~D.}\ \bibnamefont {Gunton}},\ and\
  \bibinfo {author} {\bibfnamefont {M.}~\bibnamefont {Muthukumar}},\ }\bibfield
   {title} {\bibinfo {title} {Dynamics of phase separation in a binary polymer
  blend of critical composition},\ }\href {https://doi.org/10.1063/1.458277}
  {\bibfield  {journal} {\bibinfo  {journal} {The Journal of Chemical Physics}\
  }\textbf {\bibinfo {volume} {92}},\ \bibinfo {pages} {6899} (\bibinfo {year}
  {1990})}\BibitemShut {NoStop}%
\bibitem [{\citenamefont {Wang}\ and\ \citenamefont
  {Composto}(2000)}]{Wang2000}%
  \BibitemOpen
  \bibfield  {author} {\bibinfo {author} {\bibfnamefont {H.}~\bibnamefont
  {Wang}}\ and\ \bibinfo {author} {\bibfnamefont {R.~J.}\ \bibnamefont
  {Composto}},\ }\bibfield  {title} {\bibinfo {title} {Thin film polymer blends
  undergoing phase separation and wetting: Identification of early,
  intermediate, and late stages},\ }\href {https://doi.org/10.1063/1.1322638}
  {\bibfield  {journal} {\bibinfo  {journal} {The Journal of Chemical Physics}\
  }\textbf {\bibinfo {volume} {113}},\ \bibinfo {pages} {10386} (\bibinfo
  {year} {2000})}\BibitemShut {NoStop}%
\bibitem [{\citenamefont {Weyer}\ \emph {et~al.}(2024)\citenamefont {Weyer},
  \citenamefont {Muramatsu},\ and\ \citenamefont {Frey}}]{Weyer2024}%
  \BibitemOpen
  \bibfield  {author} {\bibinfo {author} {\bibfnamefont {H.}~\bibnamefont
  {Weyer}}, \bibinfo {author} {\bibfnamefont {D.}~\bibnamefont {Muramatsu}},\
  and\ \bibinfo {author} {\bibfnamefont {E.}~\bibnamefont {Frey}},\ }\bibfield
  {title} {\bibinfo {title} {Coarsening dynamics of chemotactic aggregates},\
  }\href@noop {} {\  (\bibinfo {year} {2024})},\ \Eprint
  {https://arxiv.org/abs/2409.20100} {arXiv:2409.20100 [cond-mat.soft]}
  \BibitemShut {NoStop}%
\bibitem [{\citenamefont {Gliott}\ \emph {et~al.}(2025)\citenamefont {Gliott},
  \citenamefont {Piekarski},\ and\ \citenamefont {Cherroret}}]{Gliott2025}%
  \BibitemOpen
  \bibfield  {author} {\bibinfo {author} {\bibfnamefont {E.}~\bibnamefont
  {Gliott}}, \bibinfo {author} {\bibfnamefont {C.}~\bibnamefont {Piekarski}},\
  and\ \bibinfo {author} {\bibfnamefont {N.}~\bibnamefont {Cherroret}},\
  }\bibfield  {title} {\bibinfo {title} {Coarsening of binary bose superfluids:
  An effective theory},\ }\href {https://doi.org/10.1103/2rqw-prly} {\bibfield
  {journal} {\bibinfo  {journal} {Phys. Rev. Res.}\ }\textbf {\bibinfo {volume}
  {7}},\ \bibinfo {pages} {033189} (\bibinfo {year} {2025})}\BibitemShut
  {NoStop}%
\bibitem [{\citenamefont {Lifshitz}\ and\ \citenamefont
  {Slyozov}(1961)}]{Lifshitz1961}%
  \BibitemOpen
  \bibfield  {author} {\bibinfo {author} {\bibfnamefont {I.~M.}\ \bibnamefont
  {Lifshitz}}\ and\ \bibinfo {author} {\bibfnamefont {V.~V.}\ \bibnamefont
  {Slyozov}},\ }\bibfield  {title} {\bibinfo {title} {The kinetics of
  precipitation from supersaturated solid solutions},\ }\href
  {https://doi.org/10.1016/0022-3697(61)90054-3} {\bibfield  {journal}
  {\bibinfo  {journal} {J. Phys. Chem. Solids}\ }\textbf {\bibinfo {volume}
  {19}},\ \bibinfo {pages} {35} (\bibinfo {year} {1961})}\BibitemShut {NoStop}%
\bibitem [{\citenamefont {Wagner}(1961)}]{Wagner1961}%
  \BibitemOpen
  \bibfield  {author} {\bibinfo {author} {\bibfnamefont {C.}~\bibnamefont
  {Wagner}},\ }\bibfield  {title} {\bibinfo {title} {Theorie der alterung von
  niederschl\"agen durch uml\"osen (ostwald-reifung)},\ }\href
  {https://doi.org/10.1002/bbpc.19610650704} {\bibfield  {journal} {\bibinfo
  {journal} {Z. Elektrochem.}\ }\textbf {\bibinfo {volume} {65}},\ \bibinfo
  {pages} {581} (\bibinfo {year} {1961})}\BibitemShut {NoStop}%
\bibitem [{\citenamefont {Heermann}\ \emph {et~al.}(1996)\citenamefont
  {Heermann}, \citenamefont {Yixue},\ and\ \citenamefont
  {Binder}}]{Heermann1996}%
  \BibitemOpen
  \bibfield  {author} {\bibinfo {author} {\bibfnamefont {D.~W.}\ \bibnamefont
  {Heermann}}, \bibinfo {author} {\bibfnamefont {L.}~\bibnamefont {Yixue}},\
  and\ \bibinfo {author} {\bibfnamefont {K.}~\bibnamefont {Binder}},\
  }\bibfield  {title} {\bibinfo {title} {Scaling solutions and finite-size
  effects in the lifshitz-slyozov theory},\ }\href
  {https://doi.org/10.1016/0378-4371(96)00110-0} {\bibfield  {journal}
  {\bibinfo  {journal} {Physica A}\ }\textbf {\bibinfo {volume} {230}},\
  \bibinfo {pages} {132} (\bibinfo {year} {1996})}\BibitemShut {NoStop}%
\bibitem [{\citenamefont {Bray}\ and\ \citenamefont {Emmott}(1995)}]{Bray1995}%
  \BibitemOpen
  \bibfield  {author} {\bibinfo {author} {\bibfnamefont {A.~J.}\ \bibnamefont
  {Bray}}\ and\ \bibinfo {author} {\bibfnamefont {C.~L.}\ \bibnamefont
  {Emmott}},\ }\bibfield  {title} {\bibinfo {title} {Lifshitz-slyozov scaling
  for late-stage coarsening with an order-parameter-dependent mobility},\
  }\href {https://doi.org/10.1103/PhysRevB.52.R685} {\bibfield  {journal}
  {\bibinfo  {journal} {Phys. Rev. B}\ }\textbf {\bibinfo {volume} {52}},\
  \bibinfo {pages} {R685} (\bibinfo {year} {1995})}\BibitemShut {NoStop}%
\bibitem [{\citenamefont {Lee}\ and\ \citenamefont
  {Rutenberg}(1997)}]{Lee1997}%
  \BibitemOpen
  \bibfield  {author} {\bibinfo {author} {\bibfnamefont {B.~P.}\ \bibnamefont
  {Lee}}\ and\ \bibinfo {author} {\bibfnamefont {A.~D.}\ \bibnamefont
  {Rutenberg}},\ }\bibfield  {title} {\bibinfo {title} {Persistence, poisoning,
  and autocorrelations in dilute coarsening},\ }\href
  {https://doi.org/10.1103/PhysRevLett.79.4842} {\bibfield  {journal} {\bibinfo
   {journal} {Phys. Rev. Lett.}\ }\textbf {\bibinfo {volume} {79}},\ \bibinfo
  {pages} {4842} (\bibinfo {year} {1997})}\BibitemShut {NoStop}%
\bibitem [{\citenamefont {Soriano}\ \emph {et~al.}(2009)\citenamefont
  {Soriano}, \citenamefont {Braslavsky}, \citenamefont {Xu}, \citenamefont
  {Krichevsky},\ and\ \citenamefont {Stavans}}]{Soriano2009}%
  \BibitemOpen
  \bibfield  {author} {\bibinfo {author} {\bibfnamefont {J.}~\bibnamefont
  {Soriano}}, \bibinfo {author} {\bibfnamefont {I.}~\bibnamefont {Braslavsky}},
  \bibinfo {author} {\bibfnamefont {D.}~\bibnamefont {Xu}}, \bibinfo {author}
  {\bibfnamefont {O.}~\bibnamefont {Krichevsky}},\ and\ \bibinfo {author}
  {\bibfnamefont {J.}~\bibnamefont {Stavans}},\ }\bibfield  {title} {\bibinfo
  {title} {Universality of persistence exponents in two-dimensional ostwald
  ripening},\ }\href {https://doi.org/10.1103/PhysRevLett.103.226101}
  {\bibfield  {journal} {\bibinfo  {journal} {Phys. Rev. Lett.}\ }\textbf
  {\bibinfo {volume} {103}},\ \bibinfo {pages} {226101} (\bibinfo {year}
  {2009})}\BibitemShut {NoStop}%
\bibitem [{\citenamefont {Siegert}(1990)}]{Siegert1990}%
  \BibitemOpen
  \bibfield  {author} {\bibinfo {author} {\bibfnamefont {M.}~\bibnamefont
  {Siegert}},\ }\bibfield  {title} {\bibinfo {title} {Growth dynamics with an
  anisotropic surface tension},\ }\href
  {https://doi.org/10.1103/PhysRevA.42.6268} {\bibfield  {journal} {\bibinfo
  {journal} {Phys. Rev. A}\ }\textbf {\bibinfo {volume} {42}},\ \bibinfo
  {pages} {6268} (\bibinfo {year} {1990})}\BibitemShut {NoStop}%
\bibitem [{\citenamefont {McFadden}\ \emph {et~al.}(1993)\citenamefont
  {McFadden}, \citenamefont {Wheeler}, \citenamefont {Braun}, \citenamefont
  {Coriell},\ and\ \citenamefont {Sekerka}}]{McFadden1993}%
  \BibitemOpen
  \bibfield  {author} {\bibinfo {author} {\bibfnamefont {G.~B.}\ \bibnamefont
  {McFadden}}, \bibinfo {author} {\bibfnamefont {A.~A.}\ \bibnamefont
  {Wheeler}}, \bibinfo {author} {\bibfnamefont {R.~J.}\ \bibnamefont {Braun}},
  \bibinfo {author} {\bibfnamefont {S.~R.}\ \bibnamefont {Coriell}},\ and\
  \bibinfo {author} {\bibfnamefont {R.~F.}\ \bibnamefont {Sekerka}},\
  }\bibfield  {title} {\bibinfo {title} {Phase-field models for anisotropic
  interfaces},\ }\href {https://doi.org/10.1103/PhysRevE.48.2016} {\bibfield
  {journal} {\bibinfo  {journal} {Phys. Rev. E}\ }\textbf {\bibinfo {volume}
  {48}},\ \bibinfo {pages} {2016} (\bibinfo {year} {1993})}\BibitemShut
  {NoStop}%
\bibitem [{\citenamefont {Rutenberg}(1996)}]{Rutenberg1996}%
  \BibitemOpen
  \bibfield  {author} {\bibinfo {author} {\bibfnamefont {A.~D.}\ \bibnamefont
  {Rutenberg}},\ }\bibfield  {title} {\bibinfo {title} {Stress-free spatial
  anisotropy in phase-ordering},\ }\href
  {https://doi.org/10.1103/PhysRevE.54.R2181} {\bibfield  {journal} {\bibinfo
  {journal} {Phys. Rev. E}\ }\textbf {\bibinfo {volume} {54}},\ \bibinfo
  {pages} {R2181} (\bibinfo {year} {1996})}\BibitemShut {NoStop}%
\bibitem [{\citenamefont {Wulff}(1901)}]{Wulff1901}%
  \BibitemOpen
  \bibfield  {author} {\bibinfo {author} {\bibfnamefont {G.}~\bibnamefont
  {Wulff}},\ }\bibfield  {title} {\bibinfo {title} {On the question of speed of
  growth and dissolution of crystal surfaces},\ }\href@noop {} {\bibfield
  {journal} {\bibinfo  {journal} {Z. Kristallogr}\ }\textbf {\bibinfo {volume}
  {34}},\ \bibinfo {pages} {449} (\bibinfo {year} {1901})}\BibitemShut
  {NoStop}%
\bibitem [{\citenamefont {Herring}(1953)}]{Herring1953}%
  \BibitemOpen
  \bibfield  {author} {\bibinfo {author} {\bibfnamefont {C.}~\bibnamefont
  {Herring}},\ }in\ \href@noop {} {\emph {\bibinfo {booktitle} {Structure and
  Properties of Solid Surfaces}}},\ \bibinfo {editor} {edited by\ \bibinfo
  {editor} {\bibfnamefont {R.}~\bibnamefont {Gomer}}\ and\ \bibinfo {editor}
  {\bibfnamefont {C.~S.}\ \bibnamefont {Smith}}}\ (\bibinfo  {publisher}
  {University of Chicago Press},\ \bibinfo {address} {Chicago},\ \bibinfo
  {year} {1953})\ p.~\bibinfo {pages} {4}\BibitemShut {NoStop}%
\bibitem [{\citenamefont {Wortis}(1988)}]{Wortis1988}%
  \BibitemOpen
  \bibfield  {author} {\bibinfo {author} {\bibfnamefont {M.}~\bibnamefont
  {Wortis}},\ }\bibinfo {title} {Equilibrium crystal shapes and interfacial
  phase transitions},\ in\ \href {https://doi.org/10.1007/978-3-642-73902-6_13}
  {\emph {\bibinfo {booktitle} {Chemistry and Physics of Solid Surfaces
  VII}}},\ \bibinfo {editor} {edited by\ \bibinfo {editor} {\bibfnamefont
  {R.}~\bibnamefont {Vanselow}}\ and\ \bibinfo {editor} {\bibfnamefont
  {R.}~\bibnamefont {Howe}}}\ (\bibinfo  {publisher} {Springer Berlin
  Heidelberg},\ \bibinfo {address} {Berlin, Heidelberg},\ \bibinfo {year}
  {1988})\ pp.\ \bibinfo {pages} {367--405}\BibitemShut {NoStop}%
\bibitem [{\citenamefont {Rutenberg}\ and\ \citenamefont
  {Vollmayr-Lee}(1999)}]{Rutenberg1999}%
  \BibitemOpen
  \bibfield  {author} {\bibinfo {author} {\bibfnamefont {A.~D.}\ \bibnamefont
  {Rutenberg}}\ and\ \bibinfo {author} {\bibfnamefont {B.~P.}\ \bibnamefont
  {Vollmayr-Lee}},\ }\bibfield  {title} {\bibinfo {title} {Anisotropic
  coarsening: Grain shapes and nonuniversal persistence},\ }\href
  {https://doi.org/10.1103/PhysRevLett.83.3772} {\bibfield  {journal} {\bibinfo
   {journal} {Phys. Rev. Lett.}\ }\textbf {\bibinfo {volume} {83}},\ \bibinfo
  {pages} {3772} (\bibinfo {year} {1999})}\BibitemShut {NoStop}%
\bibitem [{\citenamefont {Johnson}(1965)}]{Johnson1965}%
  \BibitemOpen
  \bibfield  {author} {\bibinfo {author} {\bibfnamefont {C.~A.}\ \bibnamefont
  {Johnson}},\ }\bibfield  {title} {\bibinfo {title} {Generalization of the
  gibbs-thomson equation},\ }\href
  {https://doi.org/10.1016/0039-6028(65)90024-5} {\bibfield  {journal}
  {\bibinfo  {journal} {Surface Science}\ }\textbf {\bibinfo {volume} {3}},\
  \bibinfo {pages} {429–444} (\bibinfo {year} {1965})}\BibitemShut {NoStop}%
\bibitem [{\citenamefont {Mazharimousavi}\ \emph {et~al.}(2017)\citenamefont
  {Mazharimousavi}, \citenamefont {Forghani},\ and\ \citenamefont
  {Abtahi}}]{Mazharimousavi2017}%
  \BibitemOpen
  \bibfield  {author} {\bibinfo {author} {\bibfnamefont {S.~H.}\ \bibnamefont
  {Mazharimousavi}}, \bibinfo {author} {\bibfnamefont {S.~D.}\ \bibnamefont
  {Forghani}},\ and\ \bibinfo {author} {\bibfnamefont {S.~N.}\ \bibnamefont
  {Abtahi}},\ }\bibfield  {title} {\bibinfo {title} {Generalized monge gauge},\
  }\href {https://doi.org/10.1142/S0219887817500621} {\bibfield  {journal}
  {\bibinfo  {journal} {International Journal of Geometric Methods in Modern
  Physics}\ }\textbf {\bibinfo {volume} {14}},\ \bibinfo {pages} {1750062}
  (\bibinfo {year} {2017})}\BibitemShut {NoStop}%
\bibitem [{\citenamefont {Boas}(2006)}]{Boas2006}%
  \BibitemOpen
  \bibfield  {author} {\bibinfo {author} {\bibfnamefont {M.~L.}\ \bibnamefont
  {Boas}},\ }\href {https://cds.cern.ch/record/913305} {\emph {\bibinfo {title}
  {{Mathematical methods in the physical sciences}}}}\ (\bibinfo  {publisher}
  {Wiley},\ \bibinfo {address} {Hoboken, NJ},\ \bibinfo {year}
  {2006})\BibitemShut {NoStop}%
\bibitem [{\citenamefont {Jackson}(1999)}]{Jackson1999}%
  \BibitemOpen
  \bibfield  {author} {\bibinfo {author} {\bibfnamefont {J.~D.}\ \bibnamefont
  {Jackson}},\ }\href {http://cdsweb.cern.ch/record/490457} {\emph {\bibinfo
  {title} {Classical electrodynamics}}},\ \bibinfo {edition} {3rd}\ ed.\
  (\bibinfo  {publisher} {Wiley},\ \bibinfo {address} {New York, {NY}},\
  \bibinfo {year} {1999})\BibitemShut {NoStop}%
\end{thebibliography}
%

\end{document}